\documentclass{stsci_report}

\usepackage{graphicx}
\usepackage{listings}
\usepackage{xcolor}

\usepackage{siunitx}
\usepackage{todonotes}
\usepackage{pdfpages}
\usepackage{hyperref}

\usepackage{dirtytalk}
\usepackage[section]{placeins}
\usepackage{caption}
\usepackage{aas_macros}

\usepackage{float}
\usepackage{graphicx}
\usepackage{siunitx}
\usepackage{todonotes}
\usepackage{pdfpages}
\usepackage{tabularx}
\usepackage{multirow}

\usepackage{longtable}
\usepackage{caption}
\usepackage{subcaption}
\usepackage{aas_macros}
\usepackage[bottom]{footmisc}
\usepackage{wrapfig}
\usepackage{amsmath}
\usepackage{colortbl}
\usepackage{tabularray }
\usepackage{setspace} 
\usepackage{natbib}
\usepackage{listings}
\usepackage{ragged2e}

\DeclareCaptionType{equ}[][]
\usepackage{amsmath} 

\definecolor{codegreen}{rgb}{0,0.6,0}
\definecolor{codegray}{rgb}{0.5,0.5,0.5}
\definecolor{codepurple}{rgb}{0.58,0,0.82}
\definecolor{backcolour}{rgb}{0.95,0.95,0.92}

\lstdefinestyle{mystyle}{
  backgroundcolor=\color{backcolour},   
  commentstyle=\color{blue},
  keywordstyle=\color{codegreen},
  numberstyle=\tiny\color{codegray},
  stringstyle=\color{codepurple},
  basicstyle=\ttfamily\footnotesize,
  breakatwhitespace=false,         
  breaklines=true,                 
  captionpos=b,                    
  keepspaces=true,                 
  numbers=left,                    
  numbersep=5pt,                  
  showspaces=false,                
  showstringspaces=false,
  showtabs=false,                  
  tabsize=1
}

\copyrighttext{Copyright\copyright\ \the\year\ The Association of Universities for Research in Astronomy, Inc. All Rights Reserved.}

\presubtitle{Instrument Science Report WFC3 2025-04}
\title{WFC3/IR Geometric Distortion - Time Evolution of Linear Terms w.r.t. Gaia DR3}
\author{Anne O'Connor, Varun Bajaj}
\date{\today}

\begin{document}

\maketitle
\begin{Center}
\abstract{We examine the relative offsets of the linear terms in the geometric distortion between WFC3/IR and the Gaia DR3 catalog using the Mikulski Archive for Space Telescopes (MAST) pipeline WFC3/IR to Gaia DR3 alignment solutions to assess temporal stability over the lifetime of the WFC3 instrument (2009-2024). We find a period of increased uncertainty and offsets in the rotation term between 2018 and 2021, as seen in a previous analysis of WFC3/UVIS linear geometric distortion \citep{2024wfc..rept...15O}, corresponding with a period of increased jitter. We find a similar pattern of increased uncertainty between 2018 and 2021 in the skew offsets to Gaia DR3, as well.  We find no significant linear temporal evolution in the rotation, skew, or scale offsets between the WFC3/IR IDCTAB distortion solution and Gaia DR3 over the 16-year lifetime of the WFC3 instrument; however, we do see temporal evolution in the shift offsets (the difference -in pixels- between the IDCTAB and Gaia WCS positions), which are dominated by telescope pointing inaccuracy external to the WFC3/IR geometric distortion solution.  For observers requiring high-precision astrometry, we continue to recommend that observers verify or improve image alignment using the \texttt{tweakreg}\footnote{Part of the \href{https://drizzlepac.readthedocs.io/en/latest/}{DrizzlePac} package. } routine.}
\end{Center}

\section{Introduction}
\label{sec:Intro}
\subsection{WFC3/IR Geometric Distortion}
\label{sec:geo_dist intro}

The geometric distortion of the WFC3/IR detector is represented by a 4th order polynomial, described in \cite{2009wfc..rept...34K}, and stored as a reference file (the Instrument Distortion Coefficients Table or IDCTAB) to be used in the WFC3 calibration pipeline (\texttt{calwf3}) and Drizzlepac, a software package used to create distortion-corrected, cosmic-ray cleaned, and combined images\footnote{See the \href{https://drizzlepac.readthedocs.io/en/latest/}{DrizzlePac documentation} for more.}.  
Characterizing and correcting the geometric distortion of the WFC3/IR detector is crucial for image alignment and combination, and for measuring accurate positions, parallaxes, and proper motions of sources in WFC3/IR images. 

Significant variation in the linear terms over time can impact the accuracy of the geometric distortion correction, resulting in blurred combined images, and degrade the under-sampled Point Spread Function (PSF) in combined images. In 2012, the linear components of WFC3/UVIS and IR geometric distortion were analyzed over a 2-year time span and determined to be stable \citep{kozhurina2012wfc3}. \cite{2018wfc..rept....9M} then analyzed the WFC3/IR time dependence of the linear geometric distortion over an 8-year time span, concluding that the WFC3/IR detector was stable with a change of less than 0.1 pixels (0.013 arcseconds) over the eight year time span. 

This report utilizes a novel method of calculating the time evolution of WFC3/IR linear distortion terms with respect to the Gaia catalog, and presents the results over the entire lifetime of WFC3 (2009-2024).

\section{Data and Methodology}
\label{sec:Data}
\subsection{WFC3/IR Data Sets For Testing}
\label{sec:uvis_data}

The calibration pipeline processing for data in the Mikulski Archive for Space Telescopes (MAST) includes a step in which astrometric alignment to Gaia \citep{2016} is attempted for WFC3 (and ACS) images\footnote{See readthedocs in DrizzlePac for a description of the \href{https://drizzlepac.readthedocs.io/en/latest/mast_data_products/pipeline-astrometric-calibration-description.html}{Pipeline Astrometric Calibration.}}. When enough matches between sources in the WFC3 image and the Gaia DR3 catalog (\cite{brown2023gaia}) are found, the MAST pipeline attempts to find an alignment solution and the resulting aligned WCS is stored in the header (when successful). For our analysis, we aim to use every full-frame WFC3/IR external exposure with a MAST pipeline alignment solution (to Gaia DR3) collected between 2009 and 2024 in 5 commonly used wide-band filters (F105W, F110W, F125W, F140W, and F160W). The resulting dataset includes a total of 9,150 WFC3/IR images (though many datasets will be rejected to reduce the effects of outliers and poor astrometric fits, as described in Section \ref{sec:clipping}). The sample allows us to study the time evolution of the geometric distortion across the entirety of WFC3's lifetime aboard the Hubble Space Telescope, and the quantity of images and diversity of the astronomical fields in the sample (in which sources would fall on many different parts of the detector) help mitigate possible systematic errors that could arise with a small sample size or non-uniform placement of sources on the detector. Previous analyses, in comparison, exclusively used images of a single dense field. 

\subsection{Methodology}
\label{sec:methodology}

While presenting a similar analysis to that of \cite{2024wfc..rept...15O}, which analyzes the temporal variation in WFC3/UVIS linear terms, we use a slightly different methodology in this analysis. Instead of extracting a source catalog and subsequently manually aligning images/catalogs to the Gaia DR3 catalog to compute an alignment solution, we leverage MAST's calibration pipeline, which attempts to perform an astrometric alignment for each image. 

\subsubsection{MAST Pipeline: Image Alignment}
\label{sec:MAST-alignment}

When possible, the MAST pipeline computes an astrometric correction by aligning sources in the HST images to one of three reference catalogs (including GaiaDR3) \footnote{See readthedocs in DrizzlePac for a description of the \href{https://drizzlepac.readthedocs.io/en/latest/mast_data_products/pipeline-astrometric-calibration-description.html}{Pipeline Astrometric Calibration} and other reference catalogs.}. These corrections are only computed for images with an adequate number of matched sources.  The fitting between the matched source lists (WFC3 to Gaia DR3 catalog) is performed by the TweakWCS module linearfit\footnote{Please see the \href{https://tweakwcs.readthedocs.io/en/latest/source/linearfit.html}{TweakWCS readthedocs} for more info.}. For a more in-depth discussion of how TweakWCS computes these linear components (shift, rotation, scale, and skew) of the alignment solution, see WFC3 ISR 2024-15 (\cite{2024wfc..rept...15O}).

While the WFC3/UVIS investigation performed in \cite{2024wfc..rept...15O} extracted positions of sources using PSF fitting, the MAST alignment uses an algorithm similar to DAOFind (\cite{daophot}).  This algorithm is more general-purpose, as it can be used in cases where a PSF model is not available, though is less precise than PSF fitting.  Thus, the alignment errors will be inflated (compared to the UVIS investigation) due to random errors in measured centroids.  However, as we are interested in the general evolution of the linear terms over time, these random errors are unlikely to change the results of the study significantly. Additionally, since these alignment solutions were precomputed by the MAST pipeline, we did not need to derive them manually. The availability of these alignment solutions, along with the large data volume, made them a practical choice for our analysis.

\subsubsection{Extracting Transform Components from the WCS}
\label{sec:transform-WCS}

The linear terms of the absolute transformation from pixel coordinates to sky coordinates for each image can be extracted and calculated from the WCS keywords in the fits header, as explained in \cite{2024wfc..rept...15O}.  To measure the time evolution of the linear terms, we compute the absolute transformations for both the initial WCS as generated by the WFC3 calwf3 pipeline (the \say{IDCTAB WCS}) and the updated WCS after the image has been aligned to Gaia in the MAST pipeline (the \say{Gaia-aligned WCS}).  Comparing the corresponding transformations derived from each WCS allows measurement of the errors in the IDCTAB with respect to the Gaia DR3 catalog.

To calculate the error (in degrees) of absolute rotation or skew in the IDCTAB, the calculated values for the rotation or skew from the IDCTAB WCS are subtracted from their counterparts from the Gaia-aligned WCS. The ratios of the calculated values of $x$-scale and $y$-scale from the Gaia-aligned WCS are divided by their counterparts in the IDCTAB WCS, and reflect what the true (Gaia) pixel scale is relative to that calculated in the IDCTAB. Finally, we calculate the shift between the IDCTAB and Gaia DR3 frame by projecting the \texttt{CRVAL}(the RA and Dec coordinates of the reference point) of the IDCTAB into pixel space using the Gaia-aligned WCS, and subtracting this projected position from the original \verb+CRPIX+ values (pixel coordinates of the reference point).  As the CRPIX position is used as the origin for the linear transformation, the change in CRPIX is only due to the shift of the transformation (the change due to rotation, scale and skew at the origin of the transformation is 0).  However, to maintain consistency the shift is not added to the CRPIX, but the CRVAL, which represents the right ascension and declination of the CRPIX.  Thus computing the change in CRVAL (in RA and Dec) and projecting it back into pixel coordinates (relative to CRPIX) calculates how the image is shifted.

By computing all of these transformations relatively, we remove systematics such as filter-to-filter scale and rotation differences, as well as velocity aberration scale effects.  This allows the direct measurement of the accuracy of the distortion solution.

\subsubsection{Clipping the Data}
\label{sec:clipping}

To minimize any uncertainty from images with too few source matches with the Gaia DR3 catalog, we remove any WFC3/IR exposures with a MAST pipeline alignment solution that uses less than 50 sources cross-matched to Gaia DR3, resulting in 9,150 images. Next, sigma clipping is performed on the difference (or ratio, in the case of the scale term) values between each term in the IDCTAB linear transformations and the corresponding term in the Gaia-aligned linear transformations, as follows:
\begin{itemize}
    \item When plotting and analyzing the statistics of the linear terms by filter, we perform the sigma clipping for each filter independently.
    \item When plotting and analyzing the statistics for all filters together, we perform sigma clipping on the entire dataset at once.
    \item For the rotation, scale, and skew terms, we perform two iterations of sigma clipping to within \textbf{3 standard deviations} of the mean. 
    \item For the shift term, we perform two iterations of sigma clipping to within \textbf{2 standard deviations} of the mean, as it allows us to find a better linear fit and eliminate some outliers due to poor guide star positions. 
\end{itemize}

We then performed a linear fit for each computed quantity vs time using Scipy's \texttt{linregress} routine \citep{2020SciPy-NMeth}, which uses a standard linear least-squares optimization. The slope of the fit line describes how, if at all, each term evolves over time with respect to the Gaia DR3 catalog. 

\section{Results}
\label{sec:results}

 To evaluate how WFC3/IR geometric distortion has evolved over time with respect to the Gaia catalog, we visualize the offsets between the IDCTAB transforms and the Gaia-aligned transforms for each term.  Additionally, we compute the slope of the linear fit for each, which describes the temporal evolution. Any change to the geometric distortion would result in a maximum offset at the edge of the detector. As such, we provide the change at the edge of the detector in pixels for each term on the secondary y-axis.

Note that, for the discussion of each term, we are looking at the relative difference between the WFC3/IR \texttt{calwf3} pipeline (IDCTAB) linear transformations and the Gaia-aligned MAST pipeline linear transformations, rather than the absolute value of the shift, rotation, scale, or skew terms.

\subsection{Relative Shift Offsets}
\label{sec:shift}

The 0th order constant term in the 4th order polynomial geometric distortion solution is the shift term. The shift offset between the WFC3/IR IDCTAB WCS and Gaia-aligned WCS reports the amount in $x$ and $y$ (in pixels) that the position of a source in a WFC3/IR image must be shifted to match its corresponding position in the Gaia DR3 catalog. 

\begin{figure}[!ht]
    \centering
    \includegraphics[width=18cm]{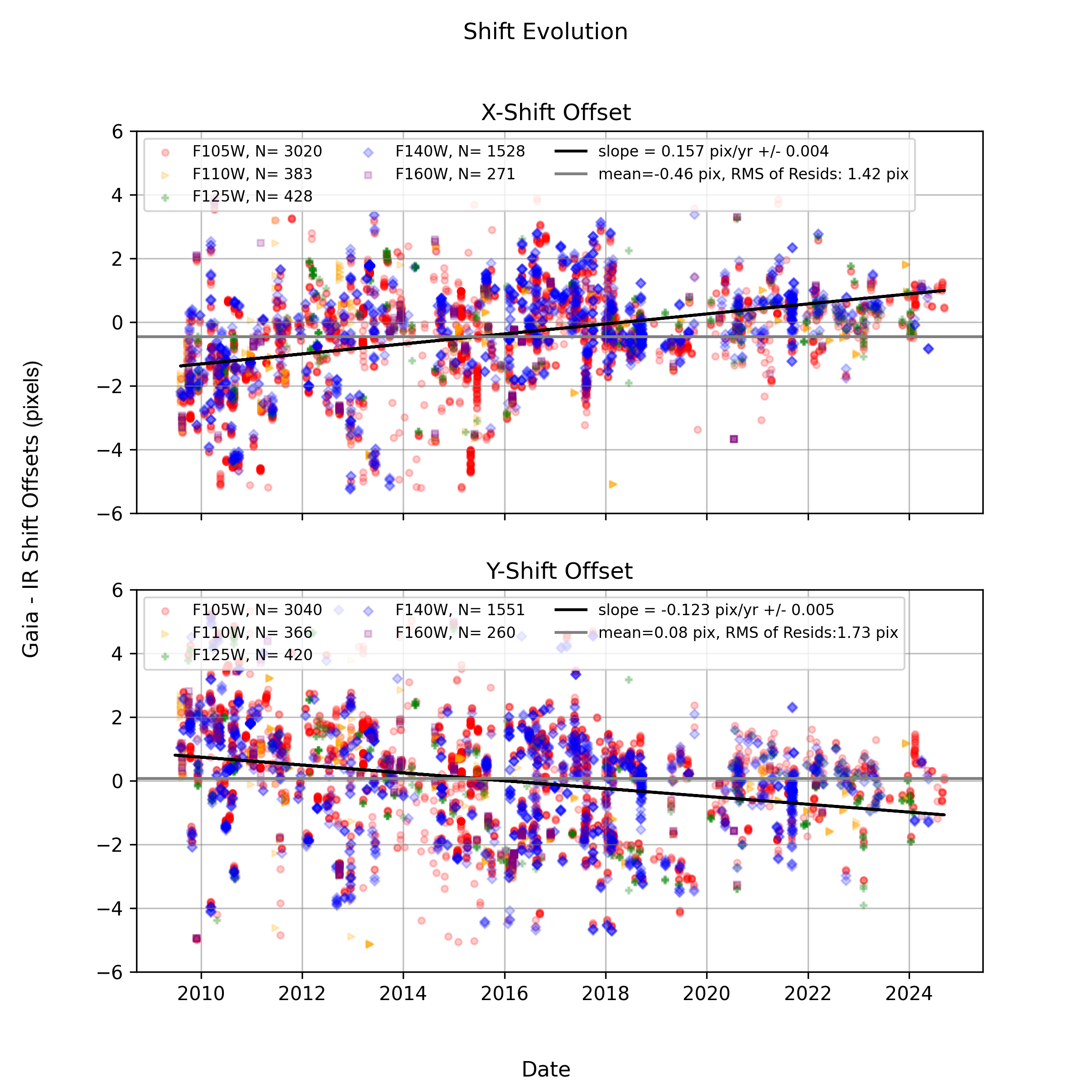}
    \caption{Shift term offsets, WFC3/IR to Gaia DR3, by filter.}
    \label{fig:shift_overplot}
\end{figure}

\begin{figure}[!ht]
    \includegraphics[width=18cm]{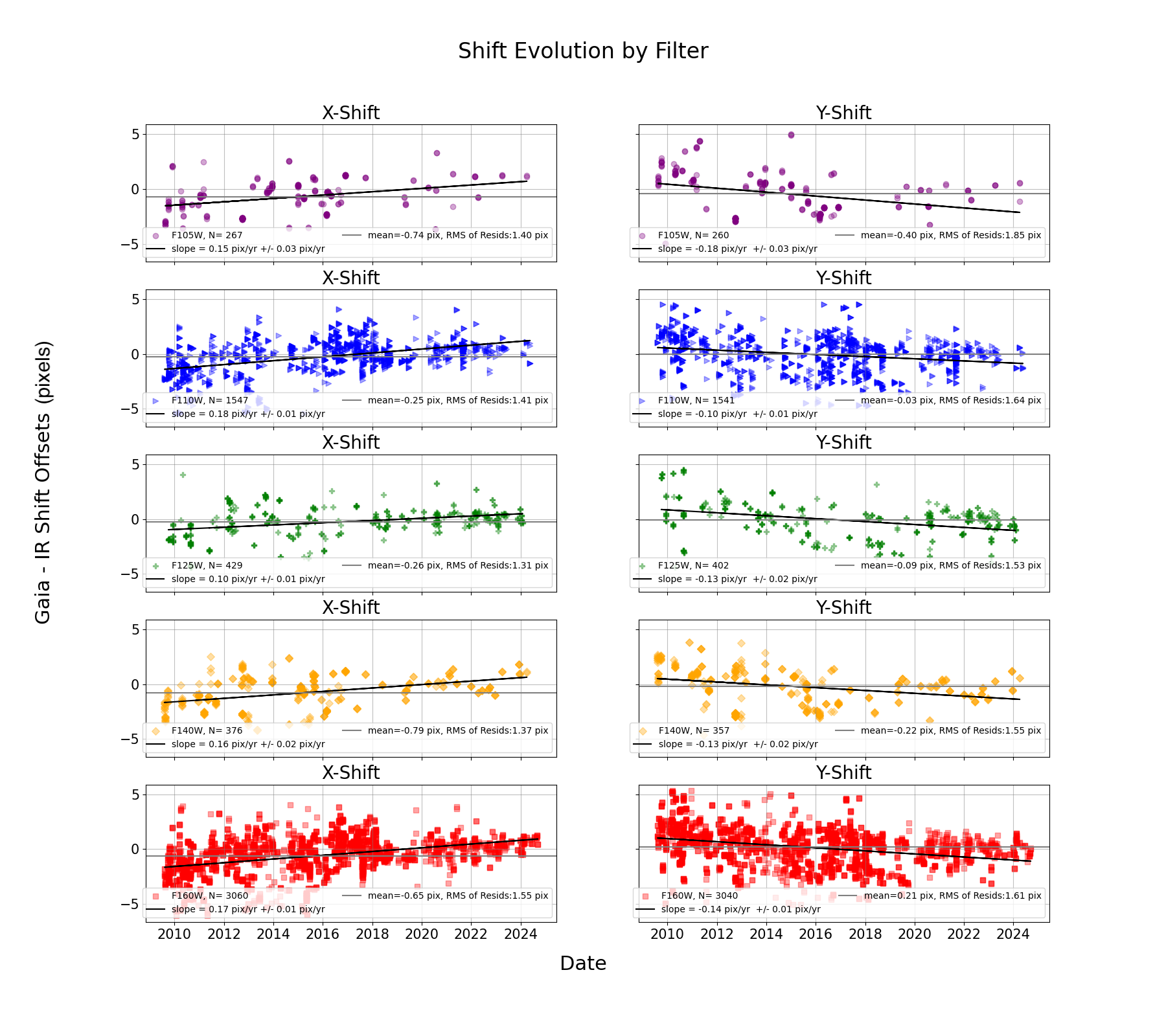}
    \caption{Shift term offsets, WFC3/IR to Gaia DR3. Each color and shape combination represents a separate filter. A line of best fit is plotted in black.}
    \label{fig:shift_subplot}
\end{figure}

The $x-$ and $y-$shift offsets between WFC3/IR and Gaia DR3 appear to evolve over time, at a rate of $0.157 \pm 0.004$ pix/yr in $x$ and $-0.123 \pm 0.005$ pix/yr in $y$, as seen in Figures \ref{fig:shift_overplot} and \ref{fig:shift_subplot}.

\clearpage

\subsection{Relative Rotation Offsets}
\label{sec:rotation}

To explore how any rotation offsets affect the positions of sources in a physical WFC3/IR image, the maximum change in pixels (at the edge of the detector) is calculated as the distance from the center of the FOV (the axis of rotation) to the edge of the detector\footnote{The WFC3/IR detector size is  $1024 \times 1024$ pixels; however, only the central 1014 pixels are light sensitive, while the 5 pixels that run around the border are reference pixels.} multiplied by the sine of the difference in rotation angle, as described by Equation \ref{eqn:rot_to_pix}. 

 \begin{equ}[h]
\captionsetup{width=.9\linewidth}
\begin{equation}
\centering
\label{eqn:rot_to_pix}
\begin{aligned}
512 \times sin(\Delta\theta)
\end{aligned}
\end{equation}
\caption*{Equation 1: used to calculate the maximum change in pixels at the edge of the detector caused by the relative rotation offset between the WFC3/IR frame and the Gaia DR3 frame.} 
\end{equ}

\begin{figure}[!b]
    \centering
    \includegraphics[width=18cm]{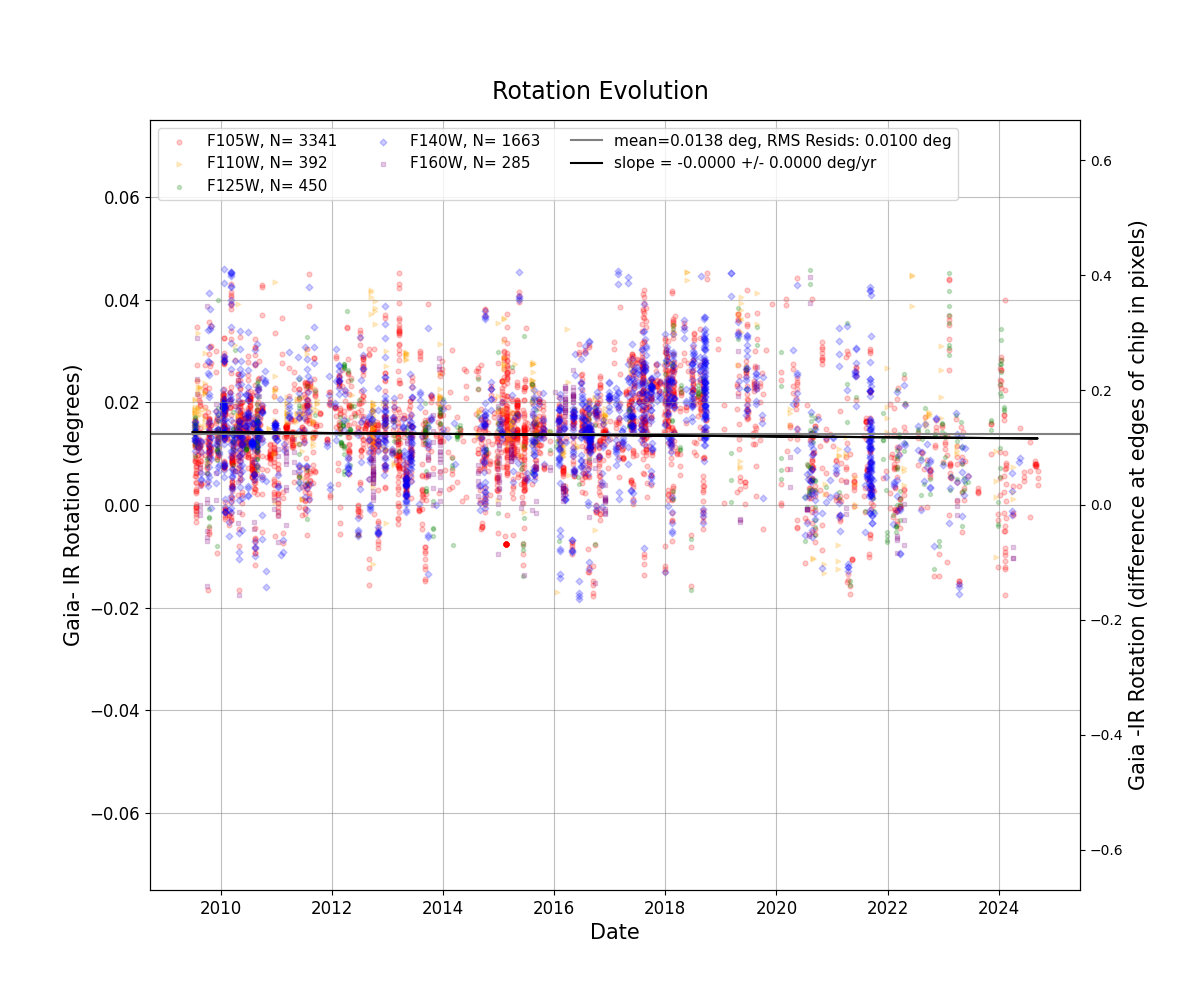}
    \caption{Rotation term offsets: WFC3/IR to Gaia DR3. Each color and shape combination represents a separate filter. A line of best fit is plotted in black.}
    \label{fig:orient_overplot}
\end{figure}

\begin{figure}[!ht]
    \centering
    \includegraphics[width=17cm]{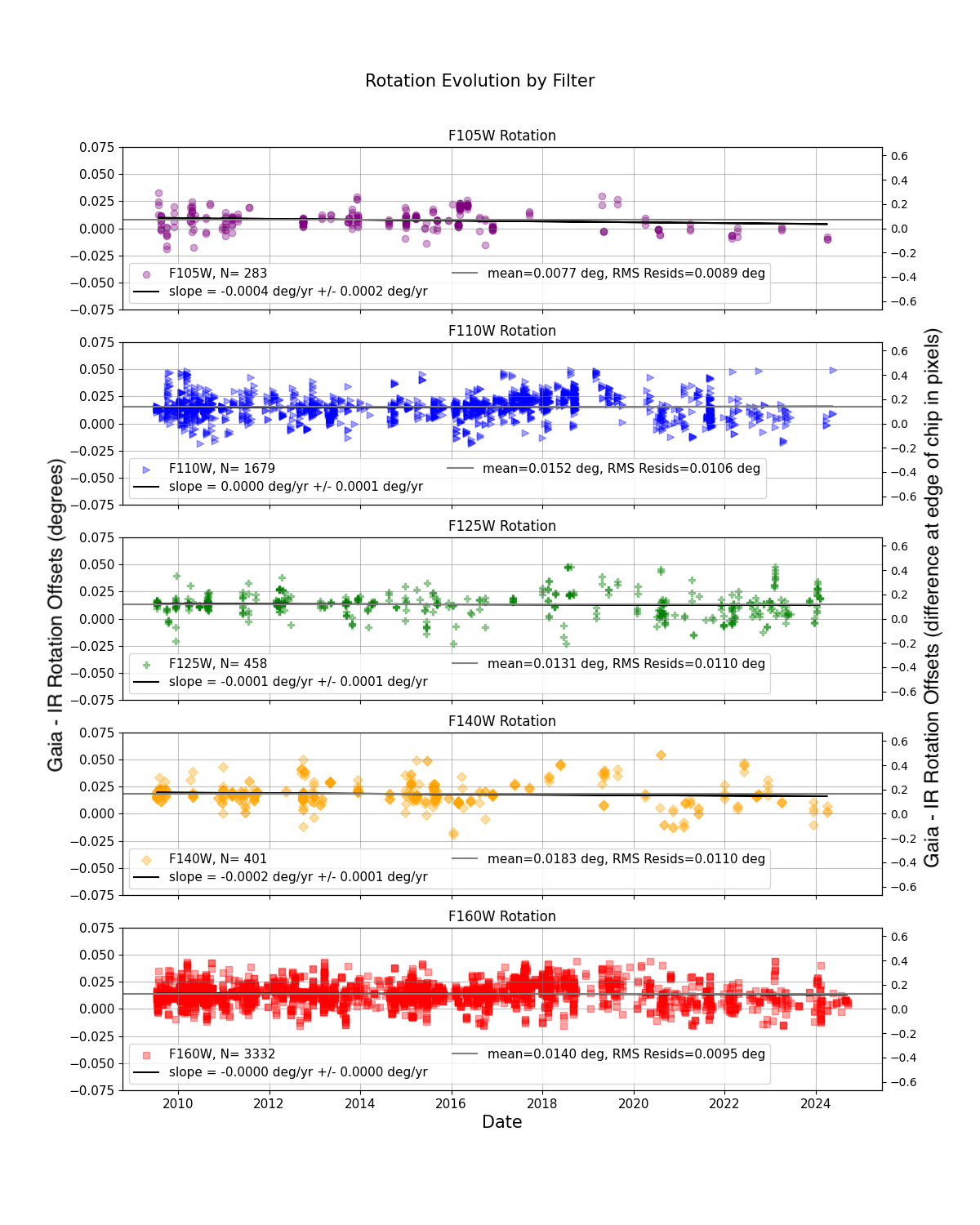}
    \caption{Rotation term offsets by filter: WFC3/IR to Gaia DR3. Each color and shape combination represents a separate filter. A line of best fit is plotted in black.}
    \label{fig:orient_subplot}
\end{figure}

The slope fit to the WFC3/IR rotation offsets with respect to Gaia DR3 over time is flat, suggesting that the rotation term has not experienced any significant \textbf{linear} temporal evolution with respect to Gaia DR3, as seen in Figures \ref{fig:orient_overplot} and \ref{fig:orient_subplot}. However, we find three distinct median rotation offsets at different epochs, meaning that the WFC3/IR geometric distortion solution has not been entirely stable over the WFC3's lifetime. The offsets in the WFC3/IR rotation to Gaia DR3 (reported as the median offset $\pm$ the RMS of the residuals) increase from a median value of $0.014 \pm 0.009$ degrees before 2018 to $0.019 \pm 0.011$ degrees between 2018 and 2021, and then jump again to $0.008 \pm 0.011$ degrees after 2021.  The timing of these jumps appears relatively consistent with what we see in the WFC3/UVIS rotation offsets (\cite{2024wfc..rept...15O}). Furthermore, the median rotation value over all time is offset from 0 by approximately 0.014 degrees (0.13 pixels at the edge), indicating a small error in the IDCTAB solution with respect to Gaia DR3.

\clearpage

\subsection{Relative Scale Offsets}
\label{sec:scale}

\begin{figure}[!t]
    \centering
    \includegraphics[width=18cm]{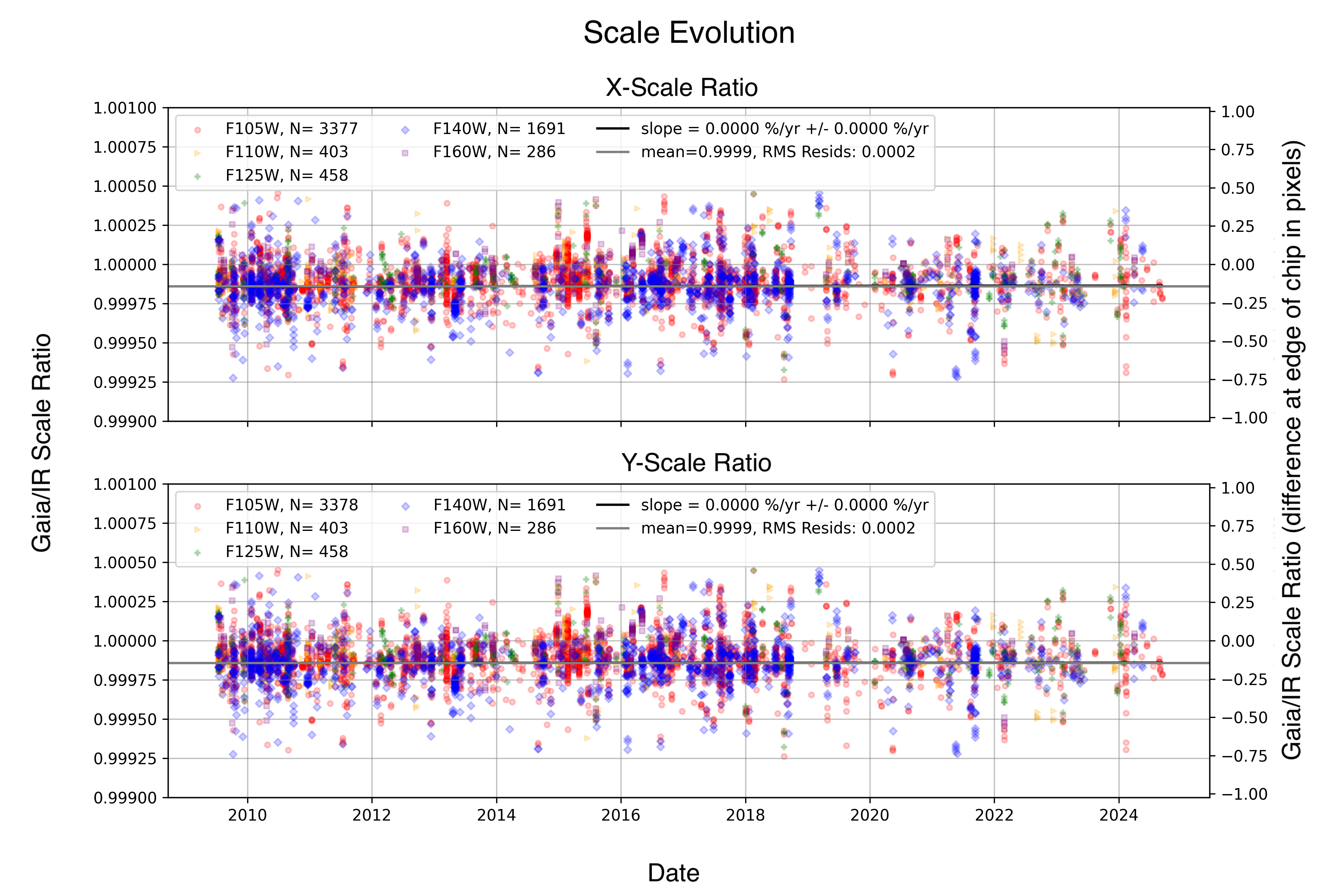}\caption    {Scale ratio: WFC3/IR to Gaia DR3. Each color and shape combination represents a separate filter. A line of best fit is plotted in black.}
    \label{fig:scale_overplot}
\end{figure}

\begin{figure}[!ht]
    \centering
    \includegraphics[width=18cm]{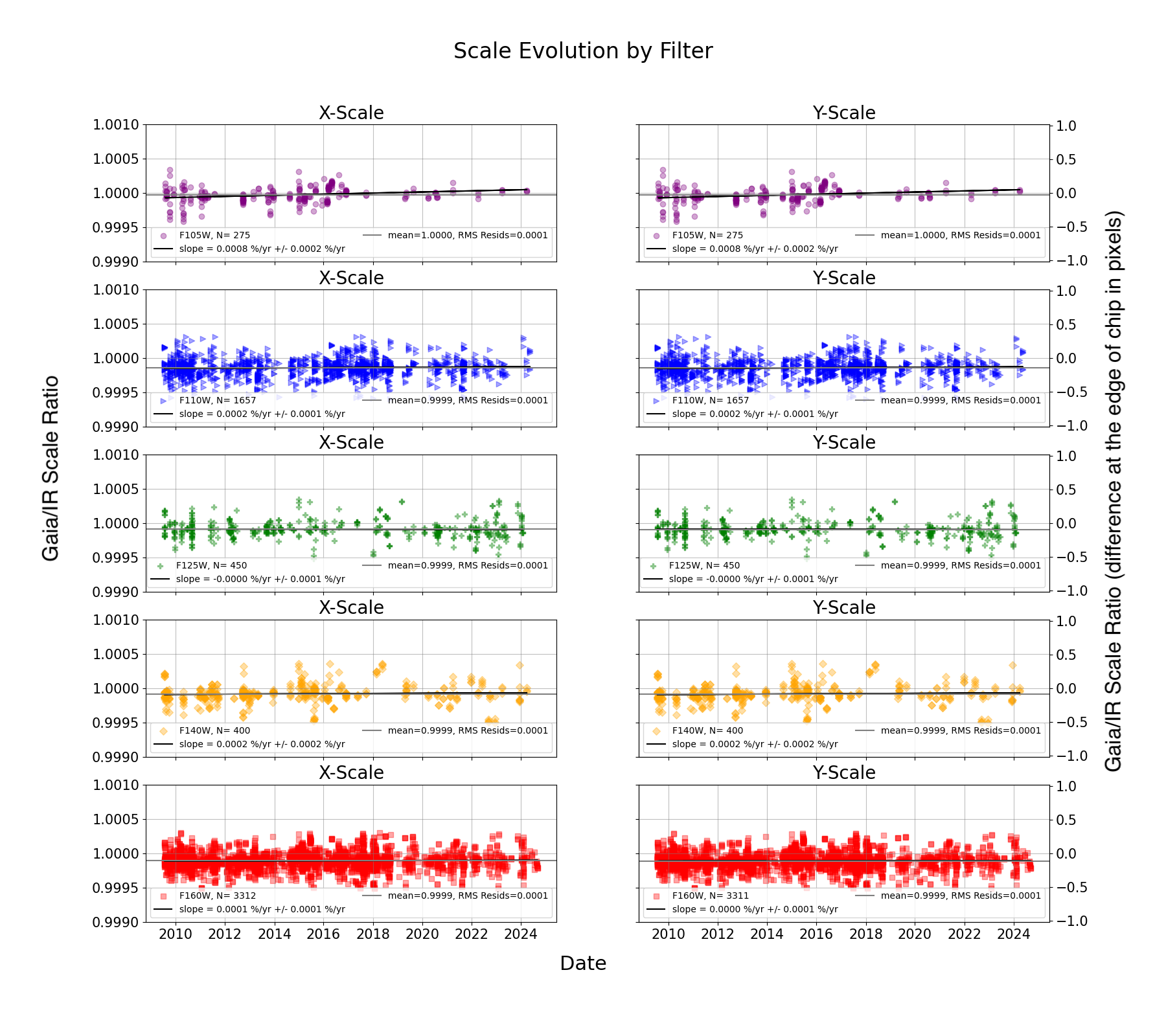}\caption    {Scale ratio by filter: WFC3/IR to Gaia DR3. Each color and shape combination represents a separate filter. A line of best fit is plotted in black.}
    \label{fig:scale_subplot}
\end{figure}

To analyze how the WFC3/IR scale has evolved over time, we compute the ratio of the Gaia $x-$scale (or $y-$scale) to the WFC3/IR $x-$scale (or $y-$scale) (see Figures \ref{fig:scale_overplot} and \ref{fig:scale_subplot}, in which a ratio of $1.0$ indicates no residual scale). To assess the greatest possible effect at the edge of the detector, in pixels, that any change in scale would have over 16 years, we compute the slope of the line of best fit for each filter (and for all data) and multiply it by the size (in pixels) of the detector, as described by Equation \ref{eqn:scale_to_pix}.  

 \begin{equ}[h]
\captionsetup{width=.9\linewidth}
\begin{equation}
\centering
\label{eqn:scale_to_pix}
\begin{aligned}
1024 \times \frac{Scale_{Gaia}}{Scale_{UVIS}} - 1024
\end{aligned}
\end{equation}
\caption*{Equation 2: used to calculate the maximum change in pixels at the edge of the detector caused by the relative scale ratio between the WFC3/IR frame and the Gaia DR3 frame.} 
\end{equ}  

Linear fits to the ratios between the WFC3/IR and Gaia $x-$ and $y-$scale terms evolve at a rate of 0\%/yr in both $x$ and $y$, suggesting that the WFC3/IR scale term has not evolved over time with respect to Gaia DR3, as seen in Figure \ref{fig:scale_overplot}.  Likewise, we do not find any statistically significant scale offsets between IR filters, in contrast to the aforementioned UVIS analysis (\cite{2024wfc..rept...15O}). As seen in Figure 
\ref{fig:scale_subplot}, there is no apparent filter-dependent offset in the scale term, with the mean $x-$ and $y-$scale ratio for each filter falling within one standard deviation of the global (all-filter) scale ratio of 0.9999.  As with the rotation term, there is a small offset from the nominal value, indicating a small error in the IDCTAB (approximately 0.12 pixels at the edge).
\clearpage

\subsection{Relative Skew Offsets}
\label{sec:skew}

The last term that we analyze is the skew term. The skew describes the total amount of non-orthogonality (difference from 90 degrees) between the principal $x$- \& $y$-axes. It is reported as an angle (in degrees), and we compute the maximum change in pixels at the edge of the detector due to a difference in skew angle by multiplying the length of the detector (1024 pixels) by the sine of the difference in skew angle, as described by Equation \ref{eqn:skew_to_pix}: 

 \begin{equ}[h]
\captionsetup{width=.9\linewidth}
\begin{equation}
\centering
\label{eqn:skew_to_pix}
\begin{aligned}
1024\times sin(\theta)
\end{aligned}
\end{equation}
\caption*{Equation 3: used to calculate the maximum change in pixels at the edge of the detector caused by the relative skew offset between the WFC3/IR frame and the Gaia DR3 frame.} 
\end{equ}  

\begin{figure}[!ht]
    \includegraphics[width=18cm]{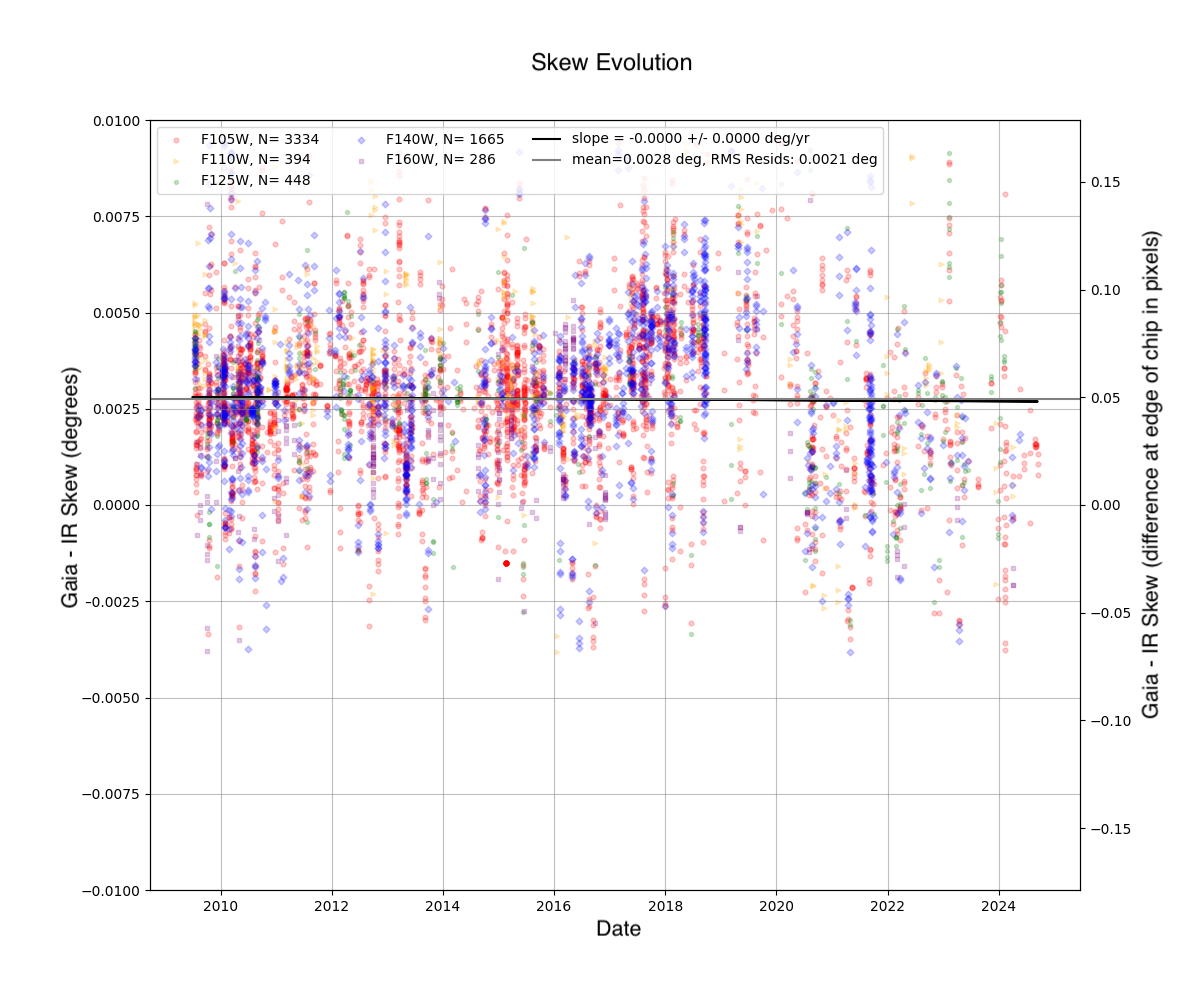}
     \caption{Skew term offsets, WFC3/IR to Gaia DR3, by filter.}
    \label{fig:skew_overplot}
\end{figure}

\begin{figure}[!ht]
    \centering
    \includegraphics[width=16cm]{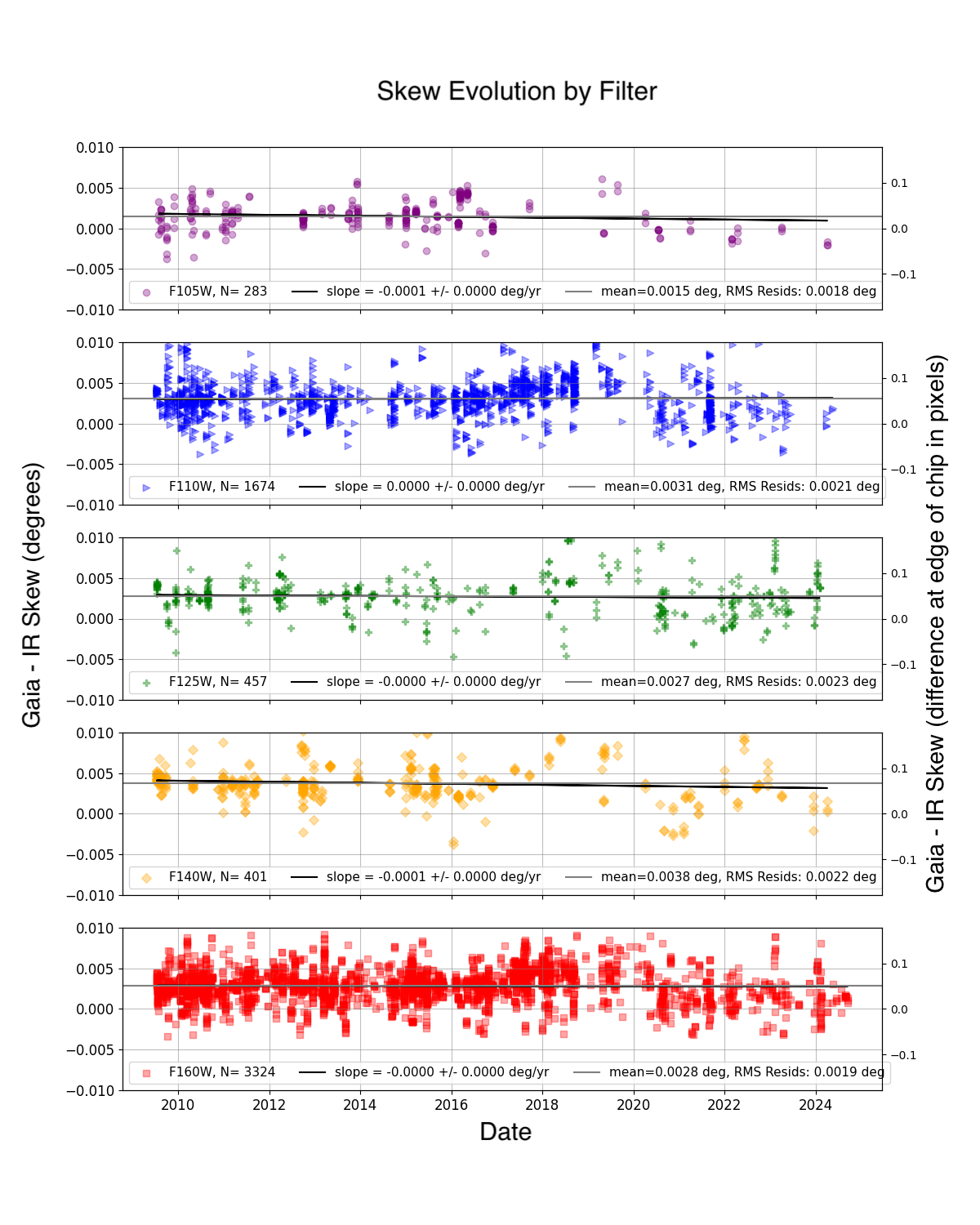}
    \caption{Skew term offsets, WFC3/IR to Gaia DR3. Each color and shape combination represents a separate filter. A line of best fit is plotted in black.}
    \label{fig:skew_subplot}
\end{figure}

The slope of the linear fit derived for the temporal evolution of the WFC3/IR skew offsets with respect to Gaia DR3 suggests no linear temporal evolution over time (a slope of 0). However, as we see with the rotation term, we find two discontinuities in the offsets in the WFC3/IR skew to Gaia DR3. We see an increase from a median value of $0.003 \pm 0.002$ degrees before 2018 to $0.004 \pm 0.002$ degrees between 2018 and 2021, and then another jump to $0.002 \pm 0.002$ degrees after 2021, as seen in Figures \ref{fig:skew_overplot} and \ref{fig:skew_subplot}. The median global offset of the skew term is 0.0028 degrees, again indicating a small error in the IDCTAB (0.05 pixels at the edge.)

\section{Discussion}
\label{sec:discussion}

The shift term is the largest term of the geometric distortion solution with respect to Gaia, and we attribute the large scatter to limitations in telescope pointing accuracy, as we do in \cite{2024wfc..rept...15O}. The scatter in the shift offsets between WFC3/IR (like WFC3/UVIS) and Gaia is largely due to sources of error outside of the distortion calibration, including uncertainty in telescope pointing. We also note that, like in \cite{2024wfc..rept...15O}, the uncertainty of the WFC3/IR to Gaia shift offsets improves after October 2017, when the Gaia astrometric catalogs were used to update the HST Guide Star Catalogs (GSC v2.40) and improve the accuracy of HST absolute astrometry \citep{hoffmann2021drizzlepac}.  For both UVIS and IR, the total shift change over time is approximately the same magnitude as the RMS scatter, making it difficult to constrain the slope with high confidence. This level of scatter is similar in both detectors (in arcseconds), though the IR channel tends to exhibit slightly more inherent scatter due to less precise centroiding. The shift term is the most weakly constrained of all components, and the slope, while small, is susceptible to bias and dominated by external factors affecting both UVIS and IR similarly. These factors likely include  drifts of the Fine Guidance Sensors and/or the WFC3 instrument in the focal plane, or perhaps changes in the positioning of the channel select mechanism (CSM) over time. 

The rotation and skew offsets, which describe how the WFC3/IR coordinate system is rotated with respect to Gaia and the non-orthogonality of the X and Y axes, respectively, do not appear to have a linear temporal evolution. However, a period of increased uncertainty and offsets in the rotation and skew terms between 2018 and 2021 is apparent, as also noted for the rotation term in the analysis of WFC3/UVIS linear geometric distortion \citep{2024wfc..rept...15O}.  Again, we note that this corresponds to a period of noticeably increased jitter in HST's pointing stability reported by \cite{2018wfc..rept....7A}. 

To check for systematic temporal variations in the linear terms that may be hidden by the large measurement uncertainties, we perform the same analyses on a subset of the F160W data with an RMS of the fit residuals  for both RA and Dec coordinates of less than 15 mas. Ultimately, we do not find any statistically significant linear temporal evolution for any term besides the shift term.

\begin{figure}[t]
    \centering
    \includegraphics[width=18cm]{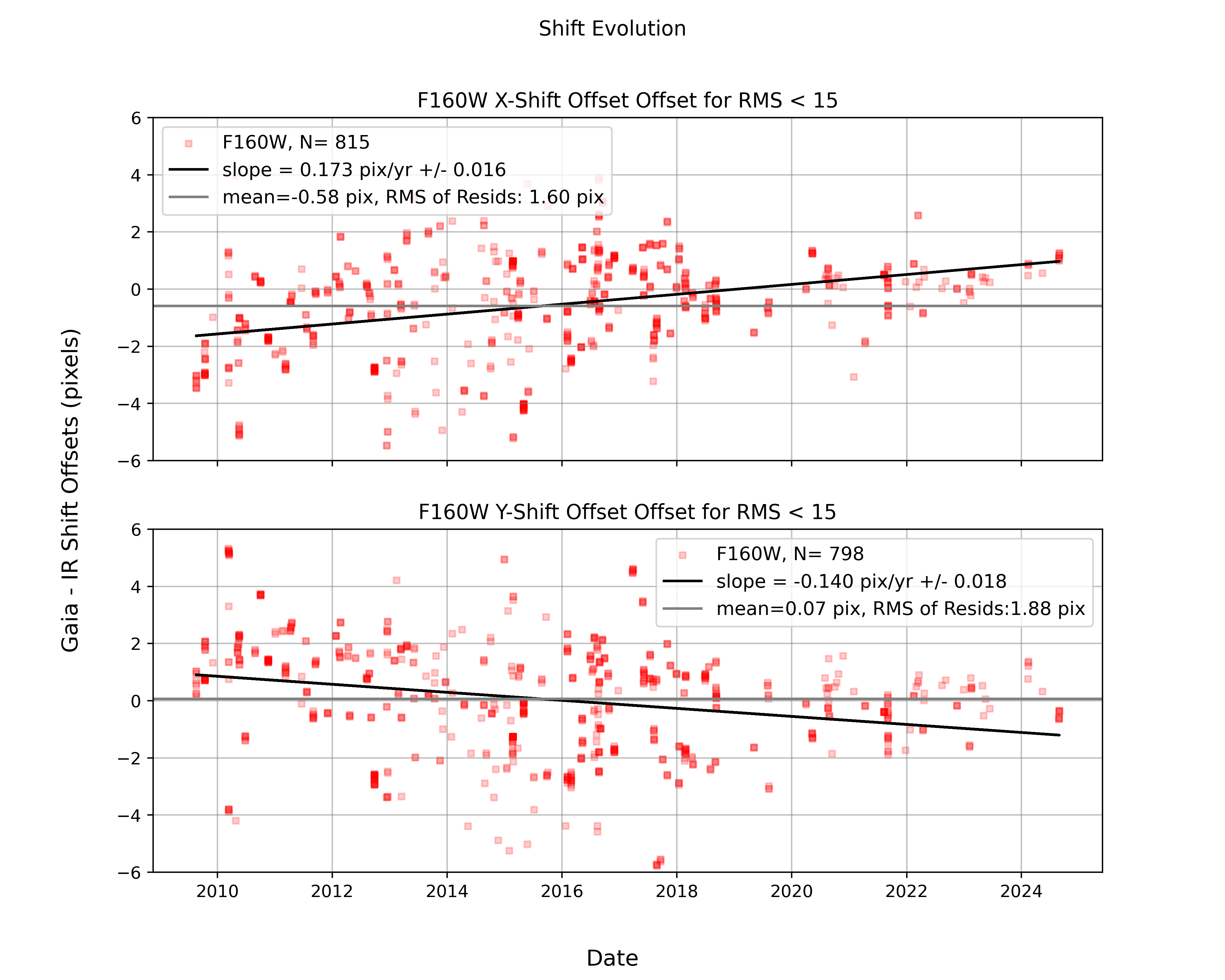}
    \caption{Shift term offsets, WFC3/IR to Gaia DR3, for F160W images with a fit RMS \textless 15 mas. A line of best fit is plotted in black.}
    \label{fig:shift_subplot_rms}
\end{figure}

\begin{figure}[!b]
    \centering
    \includegraphics[width=18cm]{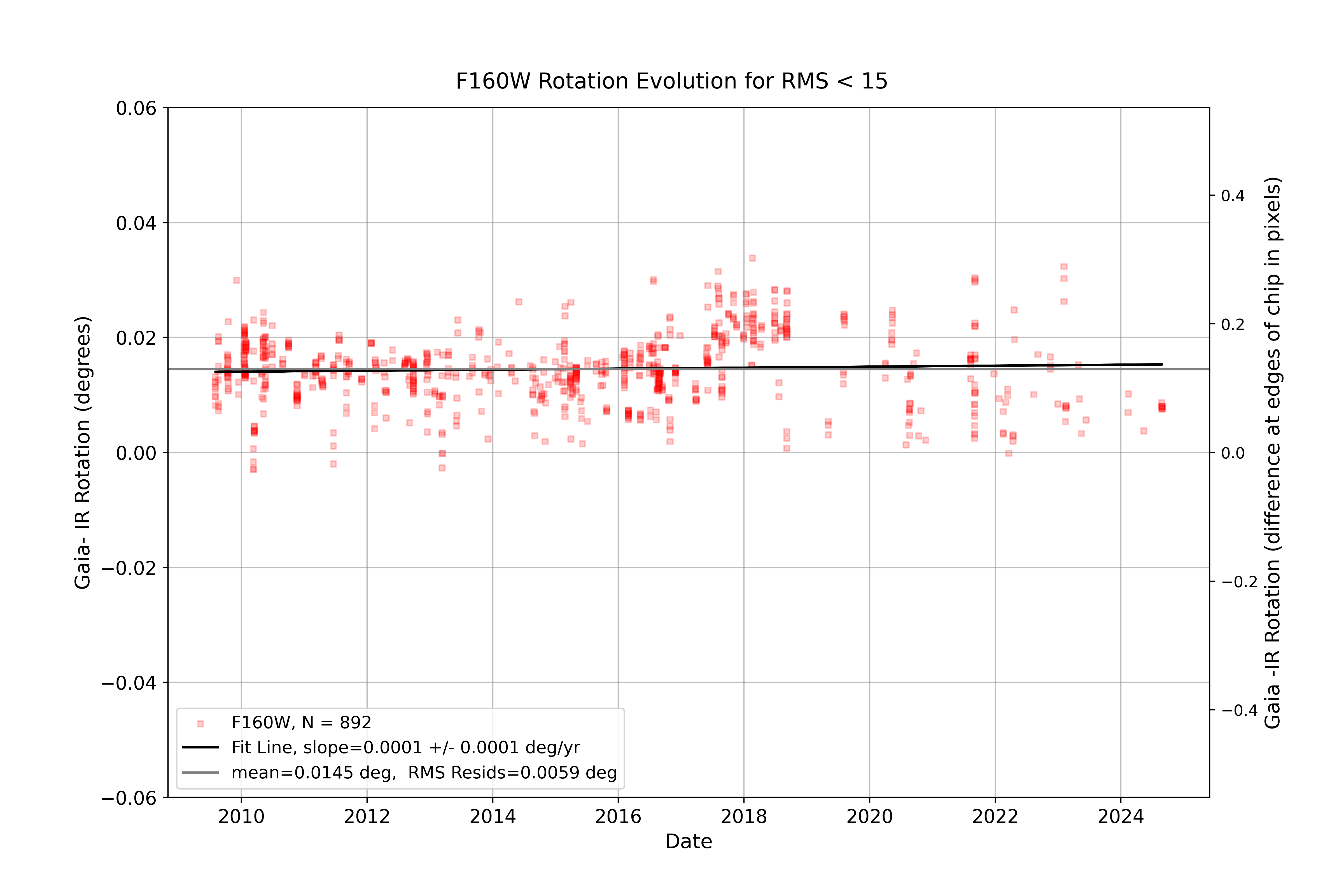}
    \caption{Rotation term offsets: WFC3/IR to Gaia DR3,  for F160W images with a fit RMS \textless 15 mas. A line of best fit is plotted in black.}
    \label{fig:orient_overplot_rms}
\end{figure}
\begin{figure}[!t]
    \centering
    \includegraphics[width=18cm]{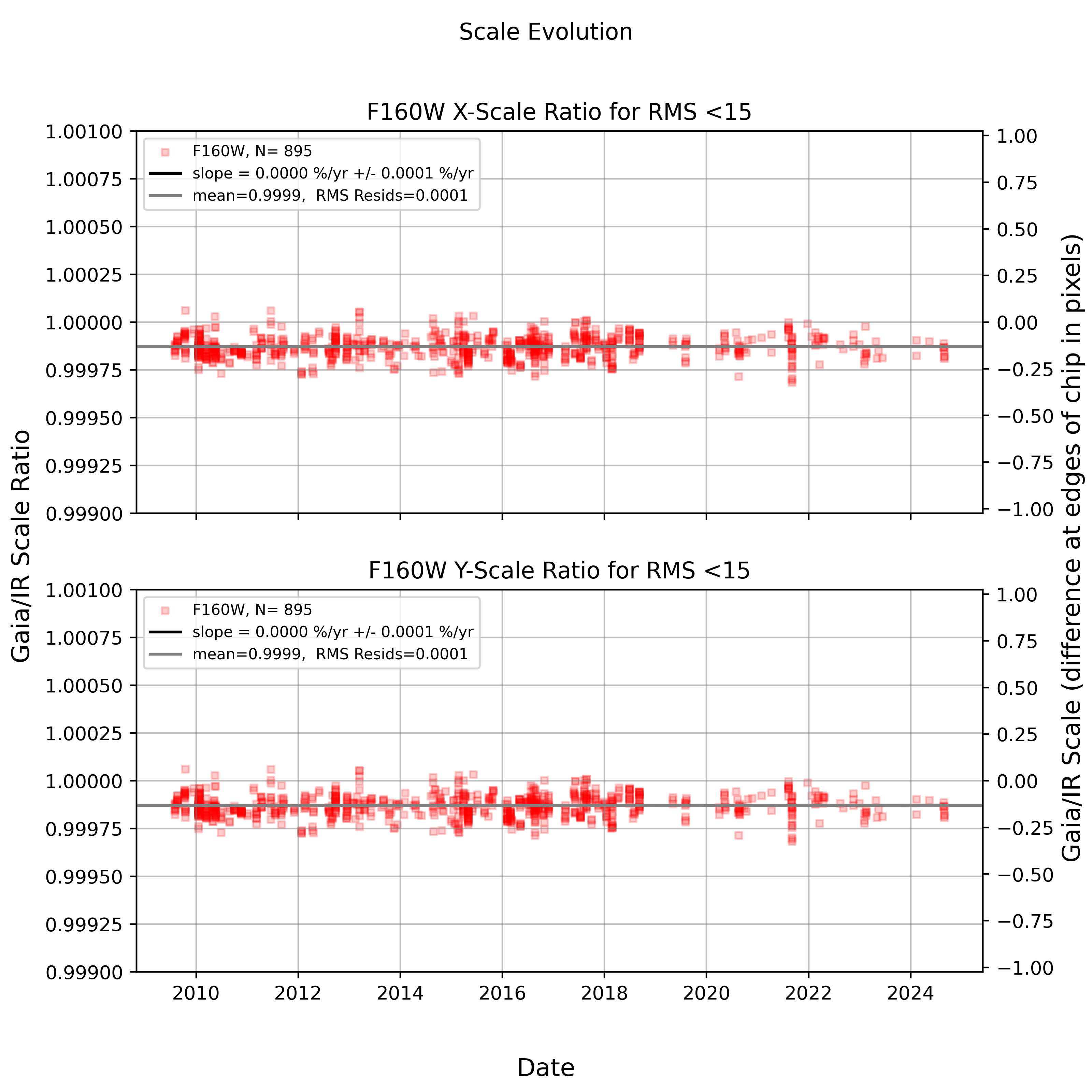}\caption    {Scale ratio: WFC3/UVIS to Gaia DR3 for F160W images with a fit RMS \textless 15 mas. A line of best fit is plotted in black.}
    \label{fig:scale_overplot_rms}
\end{figure}
\begin{figure}[!t]
    \centering
    \includegraphics[width=18cm]{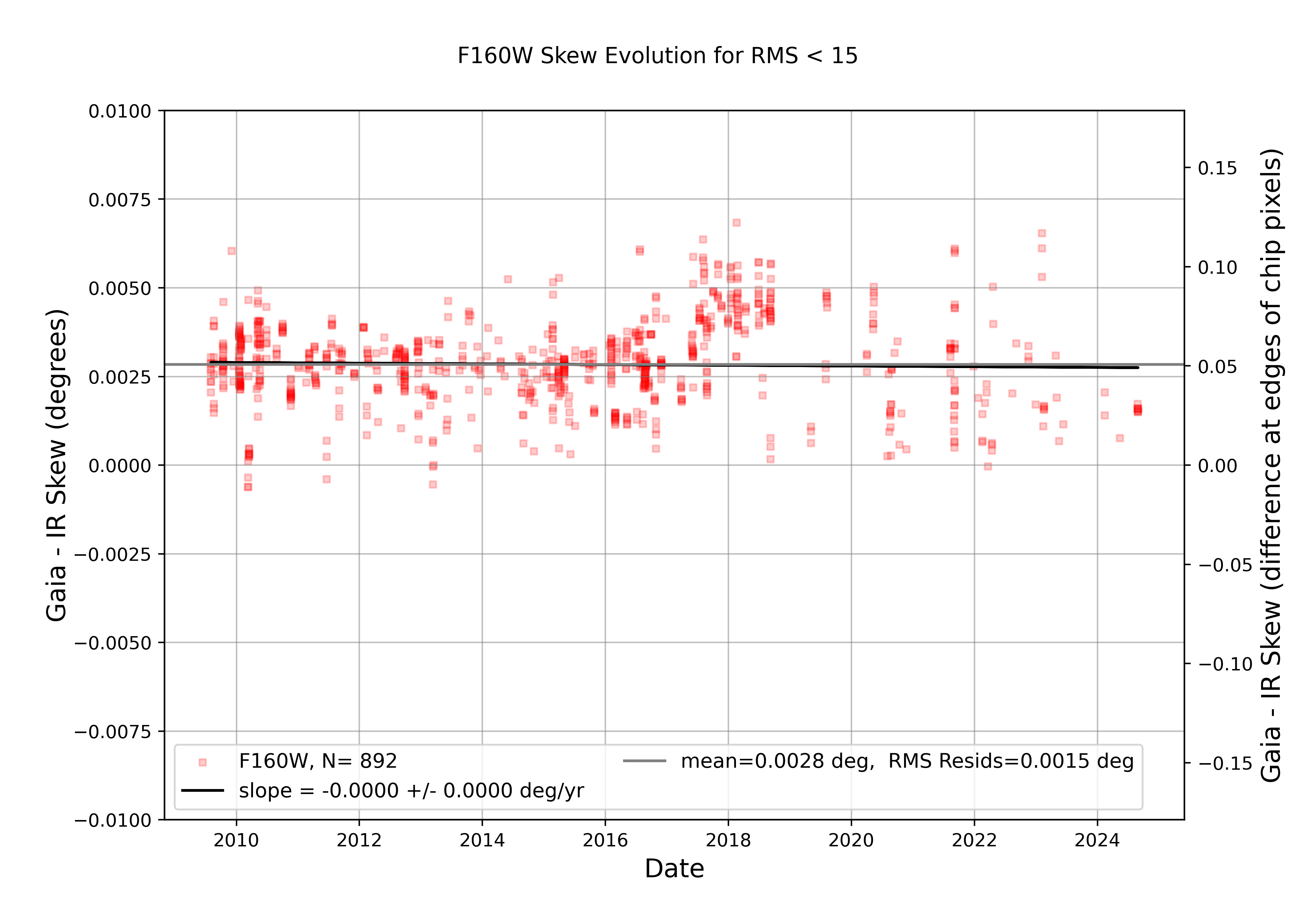}\caption    {Skew term offsets: WFC3/UVIS to Gaia DR3 for F160W images with a fit RMS \textless 15 mas. A line of best fit is plotted in black.}
    \label{fig:skew_overplot_rms}
\end{figure}
\clearpage

The presence of a small systematic offset between the IDCTAB and Gaia DR3 rotations, skews, and scales indicates that there is a small error in the astrometric calibration contained in the IDCTAB.  As Gaia catalogs were not available at the time of the initial astrometric calibration of the WFC3/IR channel, the calibration was performed using an HST-derived astrometric catalog.  This original catalog, while high precision, was not registered to a high accuracy frame (as none was available).  Thus the scale, skew, and rotation terms were not as tightly constrained, leading to small errors relative to Gaia (though the precision of the resulting distortion solution was high).  However, these accuracy errors are quite small (0.05 pixels for skew and about 0.12 pixels for scale and rotation), especially compared to the large variation in the shift offsets.   Furthermore, the scatter of these values is fairly large relative to the mean, as was the case in \citep{2024wfc..rept...15O}.  Ultimately, the larger nature of the shift terms and the scatter of the other terms cannot be fully corrected with a simple time dependent IDCTAB, and so alignment of data via TweakReg (or the MAST pipeline) is still needed for precise astrometry.  As alignment is still required to correct the shift offsets, these small systematic rotation, skew, and scale offsets are simultaneously corrected, thus invalidating the need for a corrected IDCTAB.  

\section{Conclusions}
\label{sec:Conclusions}

We find no evidence of linear temporal evolution or filter-dependent offsets in the rotation, skew, and scale terms of the WFC3/IR geometric distortion solution. The shift term, dominated by telescope pointing uncertainty, evolves at a rate of $0.157 \pm 0.004$ pix/yr in $x$ and $-0.123 \pm 0.005$ pix/yr in $y$, and the uncertainty in the shift term improves after 2017 thanks to updated HST Guide Star Catalogs. However, we do find discontinuities in the temporal evolution of the WFC3/IR rotation with respect to Gaia, similar to that observed for WFC3/UVIS \citep{2024wfc..rept...15O}. We find that the rotation offsets are stable at a value of $0.014 \pm 0.009$ degrees before 2018, jump to $0.016 \pm 0.01$ degrees between 2018 and 2021, and decrease again to $0.008 \pm 0.01$ after 2021. Similarly, the WFC3/IR skew offsets to Gaia DR3 increase from a median value of $0.003 \pm 0.002$ degrees before 2018 to $0.004 \pm 0.002$ degrees between 2018 and 2021, and then another jump to $0.002 \pm 0.02$ after 2021. However, we do not find any filter-to-filter differences in the WFC3/IR scale ratios to Gaia DR3, unlike those seen for the WFC3/UVIS scale term. The small offsets in the mean of the scale, skew, and rotation terms demonstrate small errors in the IDCTAB solution, but are fairly negligible as precise astrometry still requires alignment with TweakReg (which would correct the small errors in a given image).

\pagebreak 
We conclude that, given the stability of the WFC3/IR geometric distortion solution, the current Instrument Distortion Coefficient Table (IDCTAB) reference file in the calwf3 pipeline does not require a time-dependent solution. The shift term, while appearing to be a less stable part of the WFC3/IR geometric distortion solution, is dominated by sources of error outside of the distortion solution. We therefore suggest that observers continue to manually align their data using TweakReg\footnote{Please see the \href{https://drizzlepac.readthedocs.io/en/deployment/tweakreg.html}{TweakReg readthedocs} for more information.}, and check for an exisiting MAST pipeline alignment solution that may be sufficient for their science case. Please see WFC3 ISR 2024-15 \citep{2024wfc..rept...15O} and \cite{2022wfc..rept....6M} for more details.

\section{Acknowledgements}
 We thank the WFC3 Astrometric Calibration group for their input and review of this document. We also thank Aidan Pidgeon and Joel Green for their reviews and edits.

This work has made use of data from the European Space Agency (ESA) mission
{\it Gaia} (\url{https://www.cosmos.esa.int/gaia}), processed by the {\it Gaia}
Data Processing and Analysis Consortium (DPAC,
\url{https://www.cosmos.esa.int/web/gaia/dpac/consortium}). Funding for the DPAC
has been provided by national institutions, in particular the institutions
participating in the {\it Gaia} Multilateral Agreement.

\bibliography{ref}
\bibliographystyle{aasjournal}
\nocite{*}

\end{document}